# Shift of Fermi level by substitutional impurity-atom doping in diamond and cubic- and hexagonal-boron nitrides
## II. Generalized Gradient Approximation

Koji Kobashi

Shinko Research Co. Ltd., Division of Industrial Strategy Information
2-2-4 Wakino-hama Kaigan-dori, Chuo-ku, Kobe 651-0073, Japan
(E-mail: paris16eme2001@ybb.ne.jp)

Abstract

In succession to the first paper (arXiv 1406.6204v5), the impurity-atom concentrations when the Fermi levels are either at the valence band maximum (VBM) or the conduction band minimum (CBM) were identified for diamond, cubic boron nitride (cBN), and hexagonal boron nitride (hBN) using the Korringa-Kohn-Rostoker (KKR) scheme using the local density approximation (LDA). In the present paper, the generalized gradient approximation (GGA) was used instead of the LDA for exchange-correlation. The impurity atoms were B and N for diamond, Be, Si, and C for cBN, and Be for hBN; these impurity atoms were known in the first paper to form degenerate states by increased impurity-atom concentrations. The impurity-atom concentrations when the Fermi level was located either at the VBM or the CBM were as follows: (i) the B concentration was 0.27 at.% in B-doped diamond, (ii) the N concentration was 0.25 at.% in N-doped diamond, (iii) the concentration of Be substituting B was 0.88 at.% in cBN, (iv) the concentration of Si substituting B was 0.06 at.% in cBN, (v) the concentration of C substituting B was 0.07 at.% in cBN, (vi) the concentration of C substituting N was 0.88 at.% in cBN, and (vii) the concentration of Be substituting B was 1.80 at.% in hBN. The values of (iv) and (v) were significantly smaller than the corresponding values in paper I, but it was attributed to the input parameters used in the present paper, hence it was concluded that the computed concentrations were not sensitive to the GGA used.

In the precedent paper (paper I) [1], the shift of the Fermi level in impurity-atom doped diamond, cubic boron nitride (cBN), and hexagonal boron nitride (hBN) were investigated using the Korringa-Kohn-Rostoker (KKR) scheme [2], and the impurity-atom concentrations when the Fermi level was either at the valence band maximum (VBM) or the conduction band minimum (CBM) were computed; they were considered as the *critical concentrations* ($n_0$) from semiconductor to metal. It was found that among 17 cases that have been studied [1], the Fermi level moved either below the VBM or above the CBM only for eight cases, as shown in Table 1. In the rest of the cases, the shift of the Fermi level was restricted by the localized states that existed either in the bandgap or at the valence or the conduction band edge, even at high impurity-atom concentrations ($\leq$ 10 at.%). In paper I, the local density approximation (LDA) [2] was used for exchange-correlation. It is important, however, to repeat similar computations using a different approximation for exchange-correlation to know to what degree the $n_0$-values are affected. To this end, the general gradient approximation (GGA) [2] was used for exchange-correlation in the present paper, and the computations were done only for the eight cases shown in Table 1. It was then found that the GGA generated similar results as the LDA for the critical concentrations.

The computational procedure was the same as in paper I; the experimental lattice parameters of undoped materials were used for all computations of doped materials, the impurity-atom doping was substitutional, and no local lattice relaxation around the impurity atom was taken into account. The computations were done mostly around $n_0$ as well as for the undoped states. As for the notations of doped materials, a *p*-type diamond with $x$ at.% B impurity atoms will be denoted as *p*-D(C:*x*B), and similar notations will be used for cBN and hBN in such a way as *n*-c(B:*x*Si)N, c(B:*x*C)(N:*y*C), and *p*-h(B:*x*Be)N. The symbols, *p*, *n*, *x* and *y*, however, will be often omitted. It should be noted that since the electronic bands are progressively broadened as the dopant concentration is increased, the terms of VBM and CBM will be used to express the peak positions of the highest valence band and the lowest conduction band, respectively.

For computation, the Akai-KKR code [113] was used with the Linux Gfortran compiler [3] on a desktop computer. The basic theory and the manual of the code can be found in Refs. [4, 5], respectively. The GGA for exchange-correlation is built-in in the code [6]. The number of *k*-points in the irreducible Brillouin zone (BZ) for successful computations was 4313 for diamond and cBN, and 3251 for hBN, more than twice of the numbers in paper I. The energy range for computation was 2.5 Ry (34 eV) in which the Fermi level was located at the 3/4 of the energy range in all cases. The energy mesh in the output was 0.0125 Ry (0.169 eV). For diamond and cBN, the GGA, the coherent potential approximation (CPA) [7], the atomic-sphere approximation (ASA) [8], and the empty spheres (ESs) in vacancies [9] were used. On the other hand, the ASA was not used for hBN for the reason described in paper I.

The computed results are summarized in Table 1. The notations, *i*, *p*, *and n* in the column "Type", stand for intrinsic, *p*-type, and *n*-type semiconductors, respectively. The $n_0$-values are listed in the column "Fermi level at VBM or CBM", and compared with the results of paper I. Finally, the bandgap energies, computed in the present work and paper I, and obtained in experiments, are listed in the column "Bandgap". Note that the bandgap energies shown in the present paper were the energy differences between the CBM and the VBM in the computed band diagrams. Note also that in the figures of the density of states (DOS) and the energy bands, the Fermi level is the origin of the energy axis.

The computed DOS and band diagram of undoped diamond are shown in Figs. 1 and 2, respectively. The feature of the DOS was similar to that of paper I as well as other previous works (see references quoted in paper I). The full width of the computed valence band was 21.5 eV, comparable to an experimental width, 23.5 eV [10]. In the band diagram of Fig. 2, the indirect bandgap energy between the VBM at the Γ-point and the CBM was 4.00 eV, 27% smaller than the experimental value, 5.47 eV [11]. The CBM was approximately 3/4 from the Γ- to the X-points, in agreement with paper I and previous works as well [11, 12]. The Fermi level was located at 2.47 eV above the VBM. For B-doped diamond, $n_0 = 0.27$ at.%, similar to the value, 0.3 at.% in paper I, and it was consistent with the *critical carrier concentration* of superconductor-to-insulator transition, $n_{si} = 4 \times 10^{20}$ /cm$^3$ (0.2 at.%) of Ref. [13] for (111), (100), and (110) oriented B-doped diamond films. On the other hand, for N-doped diamond, $n_0 = 0.25$ at.%, smaller than the value 0.4 at.% in paper I.

For undoped cBN, the DOS and band diagram are shown in Figs. 3 and 4, respectively. The indirect bandgap energy between the VBM at the Γ-point and the CBM at the X-point in Fig. 4 was 4.36 eV, 31% smaller than the experimental bandgap, 6.36 eV [14]. Notice that the Fermi level was located at 4.16 eV above the VBM at the Γ-point, or only 0.20 eV below the CBM, as seen in Fig. 4. For the case of *p*-c(B:Be)N, the computed $n_0 = 0.88$ at.%, similar to the value 0.9 at.% in paper I. On the other hand, for the case of *n*-c(B:Si)N, $n_0 = 0.06$ at.%, significantly smaller than 0.3 at.% in paper I. The energy difference between the CBM and the Fermi level is shown by open circles in Fig. 5 as a function of the Si impurity concentration. It is seen that the small $n_0$ value was obtained due to the unusual behavior of the Fermi shift; it moved quickly toward the CBM with $n_{Si}$, and once it was positioned above the CBM, it stayed approximately in the same position between $n_{Si} = 0.1$ and 0.3 at.%, and then moved up deeper into the conduction band. If the points of $n_{Si}$ = 0.0, 0.3, and 0.4 are linked with a smooth curve, it crosses at ~ 0.2 at.%, which is closer to the value 0.3 at.% in paper I. To further investigate the behavior of the Fermi level, additional computations were done just by replacing the GGA with the LDA that was used in paper I. As a result, the computed $n_0$ = 0.06 at.%, as shown in Table I, and the behavior of the Fermi level with $n_{Si}$ was the same as in the present paper. It therefore follows that the behavior of the Fermi level with $n_{Si}$ in Fig. 5 originated from the input parameters for the Akai-KKR code, and has nothing to do with the exchange-correlation, the GGA and the LDA.

A similar result was obtained for n-c(B:C)N, as seen in Table I. The $n_0$-value was 0.07 at.%, far smaller than 0.3 at.% obtained in paper I. Like in the case of n-c(B:Si)N described above, the Fermi level moved up to the CBM as the C impurity concentration, $n_C$, was increased, stayed approximately in the same position between $n_C$ = 0.1 and 0.3 at.%, and then moved up into the conduction band. The same argument holds in the present case as in the case of c(B:Si)N. By contrast, the present result of $n_0$ for cB(N:C) was similar to that in paper I. For the case of c(B:$x$C)(N:$y$C), the Fermi level shifted toward the VBM when y was greater than x, while it moved toward the CBM when x was greater than y.

The computed DOS and band diagram of undoped hBN are shown in Figs. 6 and 7, respectively. The DOS of Fig. 6 is very similar to that of paper I. Among many band diagrams in previous articles, the band diagram of Ref. [15] is most similar to Fig. 6. The conduction band was the lowest at the Γ-point. The valence bands facing to the bandgap were fairly flat between the A-Γ and M-L points, and in the present computation, the VBM was at the M-point. The bandgap energy was 3.536 eV, only 62% of the experimental values, 5.7 eV [16]. For h(B:Be)N, $n_0$ = 1.80 at.%, similar to the result in paper I.

In summary, the computed bandgaps using the GGA was slightly smaller than those using the LDA for undoped diamond, cBN, and hBN. The marked difference between the present paper and paper I was that the $n_0$-values for n-c(B:Si)N and n-c(B:C)N were significantly smaller in the present paper, but this originated not from the different exchange-correlation, *i.e.* the GGA *vs.* the LDA, but from other input parameters, most likely the number of *k*-points in the BZ. Finally, it should be mentioned that experimental study of semiconducting and metallic properties of C-doped cBN is of interest as its semiconducting type depends on the C concentrations at the B and the N sites.

Acknowledgement

The author thanks Hisazumi Akai for useful comments, and Jeffrey T. Glass for kind encouragement. He is particularly indebted to Hitoshi Gomi for useful discussion on applying the Akai-KKR code for covalently bonded solids consisting of light elements.

References

[1]  K. Kobashi, arXiv:cond-mat/ 1406.6204v5 (2014).

[2]  R. M. Martin, Electronic Structure (Cambridge Univ. Press, Cambridge, 2004).

[3]  Gfortran is the name of the GNU Fortran project; https://gcc.gnu.org/wiki/GFortran.

[4]  http://kkr.phys.sci.osaka-u.ac.jp.

[5]  G. H. Fecher, http://ghfecher.de/Run_AKAI-KKR.pdf.

[6]  J. P. Perdew, J. A. Chevary, S. H. Vosko, K. A. Jackson, M. R. Pederson, D. J. Singh, and C. Fiolhais, Phys. Rev. B, **46**, 6671 (1992).

[7]  P. Soven, Phys. Rev. **156**, 809 (1967).

[8]  O. K. Anderson and O. Jepson, Physica, **91B**, 317 (1977).

[9]  N. E. Christensen, Phys. Rev. B, **37**, 4528 (1988); J. Hafner, Acta Mater. **48**, 71 (2000); M. R. Salehpour and S. Satpathy, Phys. Rev. B, **41**, 3048 (1990-I).

[10] T. Yokoya, T. Nakamura, T. Matushita, T. Muro, H. Okazaki, M. Arita, K. Shimada, H. Namatame, M. Taniguchi, Y. Takano, M. Nagao, T. Takenouchi, H. Kawarada, and T. Oguchi, Sci. Technol. Adv. Mater. **7**, S12 (2006).

[11] *Diamond: Electronic Properties and Applications* (Springer, Berlin, 1995), edited by L. S. Pan and D. R. Kania.

[12] T. Kotani and H. Akai, Phys. Rev. B, **54**, 16502 (1996-I).

[13] A. Kawano, H. Ishiwata, S. Iriyama, R. Okada, T. Yamaguchi, Y. Takano, and H. Kawarada, Phys. Rev. B, **82**, 085318 (2010).

[14] D. A. Evans, A. G. McGlynn, B. M. Towlson, M. Gunn, D. Jones1, T. E. Jenkins, R. Winter and N. R. J. Poolton, J. Phys.: Condens. Matter, **20**, 075233 (2008).

[15] K. Kobayashi, http://www.bandstructure.jp/.

[16] K. Watanabe and T. Taniguchi, Int. J. Appl. Ceram. Technol. **8**, 977 (2011); B. Arnaud, S. Lebègue, P. Rabiller, and M. Alouani, Phys. Rev. Lett. **96**, 026402 (2006).

Table 1. Summary of computed results on diamond, cBN, and hBN. For the notations, see Sec. 3 in the text.

| Materials | Type | Fermi level at VBM or CBM $n_0$ (at.%) | | Bandgap (eV) | | |
|---|---|---|---|---|---|---|
| | | GGA | LDA | This work | Ref. | Exp. |
| Diamond | *i* | | | 4.00 | 4.08 | 5.47 [11] |
| D(C:B) | *p* | 0.27 | 0.3 | | | |
| D(C:N) | *n* | 0.25 | 0.4 | | | |
| Cubic BN | *i* | - | | 4.36 | 4.57 | 6.36 [14] |
| c(B:Be)N | *p* | 0.88 | 0.9 | | | |
| c(B:Si)N | *n* | 0.06 | 0.06* | | | |
| c(B:C)N | *n* | 0.07 | 0.07* | | | |
| cB(N:C) | *p* | 0.88 | 0.9 | | | |
| c(B:C)(N:C) | *p, n* | - | - | | | |
| Hexagonal BN | *i* | | | 3.60 | 3.7 | 5.7 [16] |
| h(B:Be)N | *p* | 1.80 | ~2 | | | |

*Results obtained using the LDA in the present computational scheme.

Figure Captions

FIG. 1. DOS of undoped diamond. The origin of the energy is the Fermi level.

FIG. 2. Band diagram of undoped diamond.

FIG. 3. DOS of undoped cBN. The origin of the energy is the Fermi level.

FIG. 4. Band diagram of undoped cBN.

FIG. 5. Fermi levels of *n*-c(B:Si)N, shown by open circles, and *n*-c(B:C)N, shown by closed circles, as a function of impurity-atom concentration.

FIG. 6. DOS of undoped hBN. The origin of the energy is the Fermi level.

FIG. 7. Band diagram of undoped hBN. The CBM is at the $\Gamma$-point, while the VBM is at the M-point.

Fig. 1

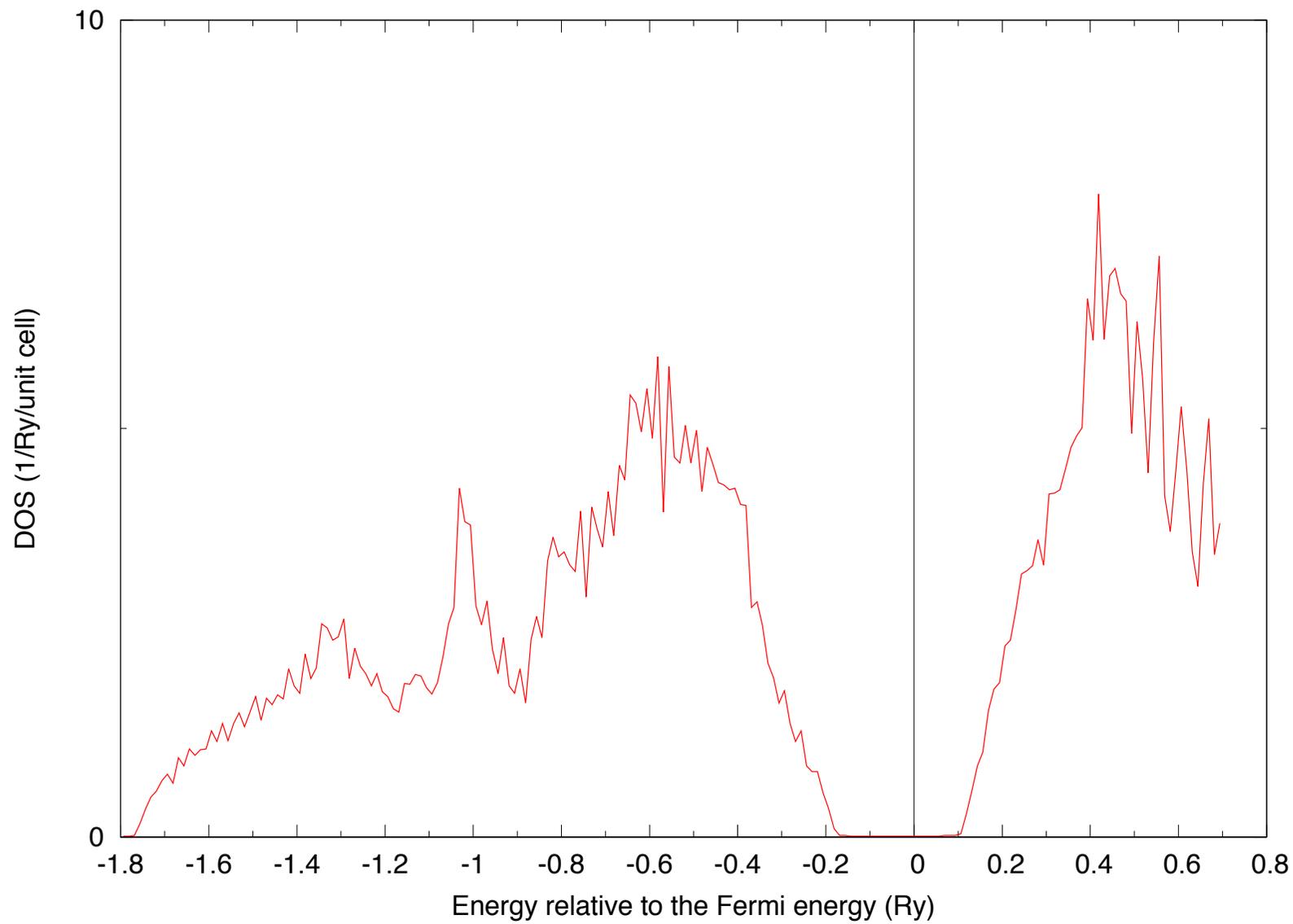

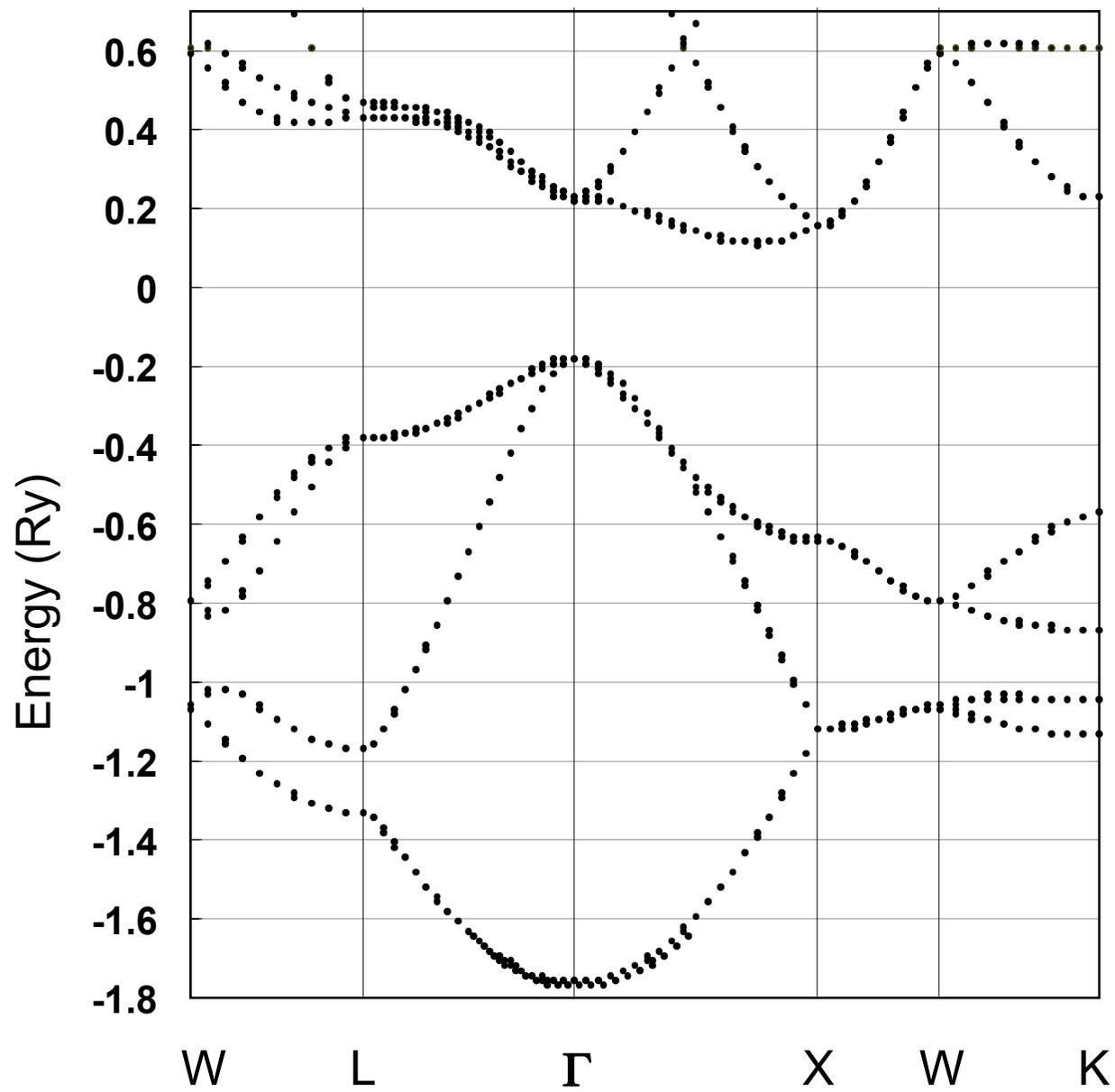

Fig. 2

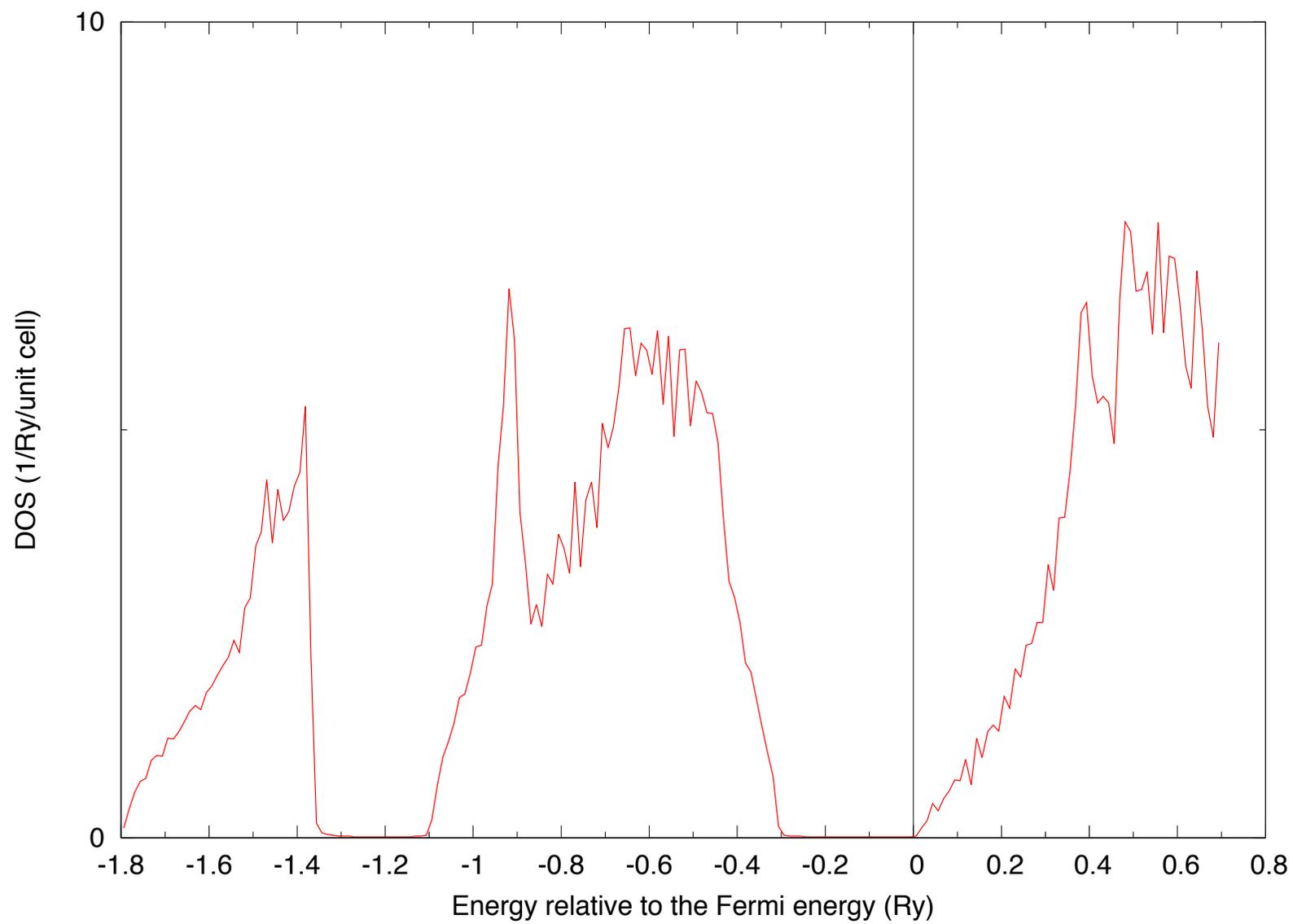

Fig. 3

Fig. 4

Energy (Ry) vs. W L Γ X W K

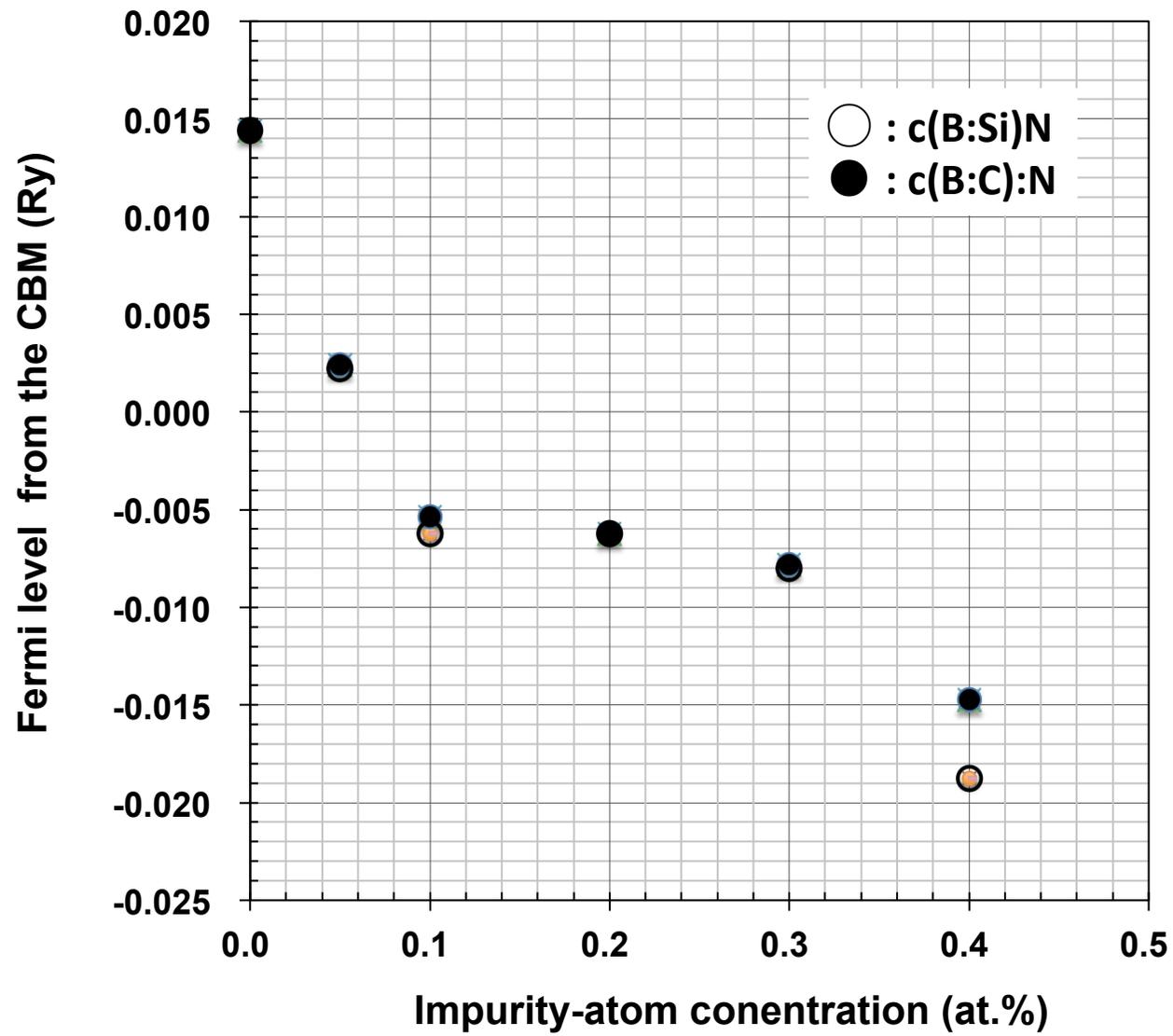

Fig. 5

Fig. 6

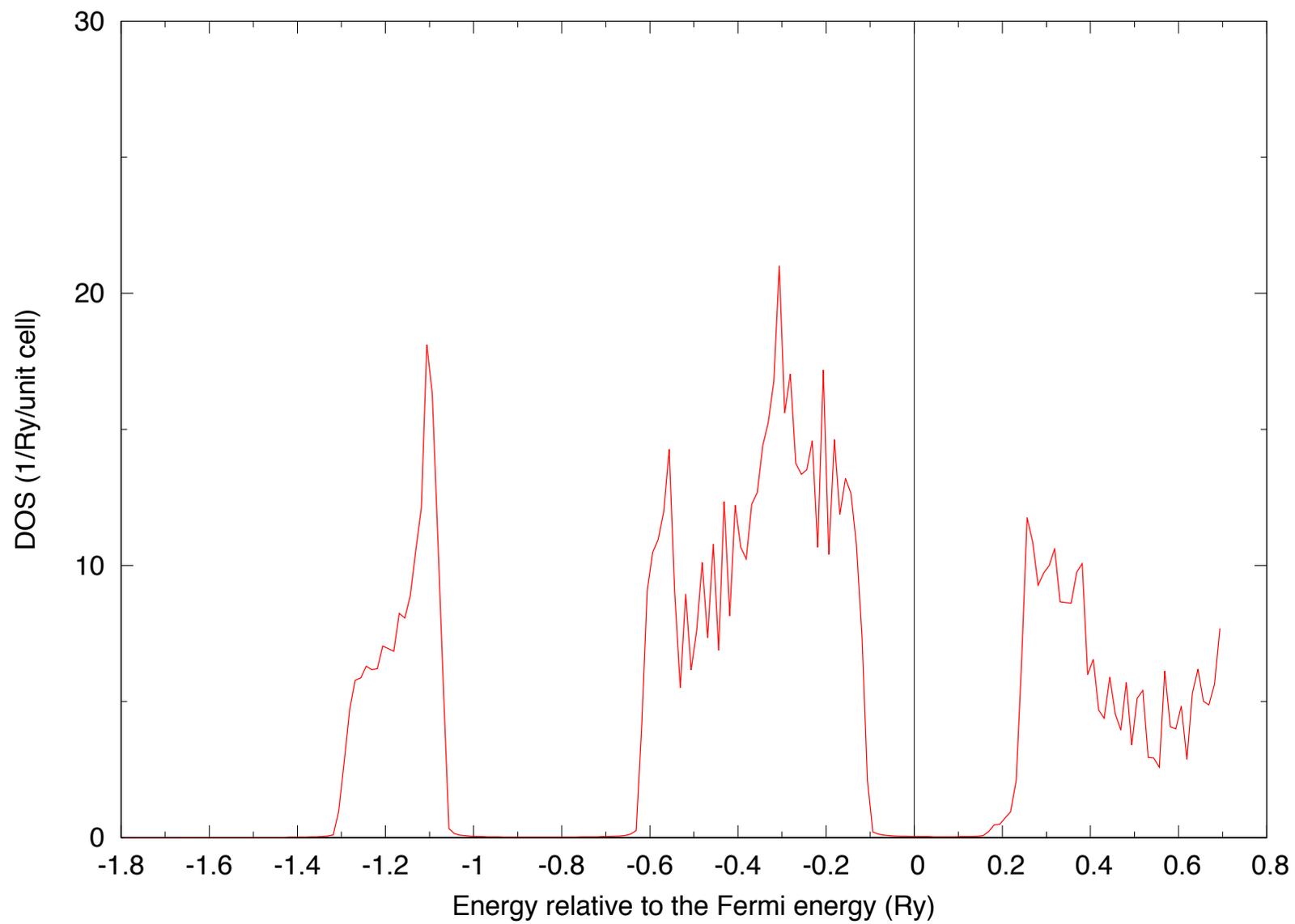

Fig. 7

Energy (Ry) vs. Γ K H A Γ M L A